\documentclass[twocolumn,showpacs]{revtex4}
\usepackage{epsf,graphicx}
\usepackage{subfigure}
\usepackage{amsmath,amssymb,amsfonts}
\begin{document}
\newcommand{\avg}[1]{\left\langle{#1}\right\rangle}
\newcommand{\davg}[1]{\left\langle\!\left\langle{#1}\right\rangle\!\right\rangle}
\newcommand{\ovl}[1]{\overline{#1}}
\renewcommand{\l}{\left}
\renewcommand{\r}{\right}
\author{A.~De~Martino, E.V.~Votyakov and D.H.E.~Gross}
\affiliation{Hahn-Meitner-Institut, Bereich Theoretische Physik,
Glienickerstr. 100, 14109 Berlin (Germany)}
\title{\sffamily Asymmetric binaries and hierarchy of states in self-gravitating systems}

\begin{abstract}
We complete the analysis of the equilibrium shapes of rotating
self-gravitating gases initiated in Phys. Rev. Lett. {\bf 89}
031101 (2002) by studying the formation and thermodynamics of
asymmetric binaries and by introducing a family of order
parameters to classify the possible equilibrium states.
\end{abstract}

\maketitle

Gravity-dominated systems such as classical self-gravitating gases
are known to possess highly non-standard thermodynamic properties
\cite{binney&tremaine87,padmanabhan90,padmata}. The long-range
nature of the Newtonian potential prevents thermodynamic
functionals from being extensive \cite{gallavotti99}, causing such
features as the inequivalence of statistical ensembles, the
occurrence of negative specific heats and of collapse transitions,
and the existence inhomogeneous ground states
\cite{lyndenbell67,lyndenbell68,thirring70,devega96,chavanis01}.
These attributes are not peculiar to gravitating gases alone.
Similar traits mark the behaviour of systems with long-range,
non-integrable potentials \cite{barre01}, of finite systems or of
systems of a size comparable to the interaction range, such as
atomic clusters or highly excited (``hot'') nuclei
\cite{gross82,gross140}, are even present in certain short-ranged
models \cite{compagner} and in general at phase separations
\cite{gross174}. Besides the relevance to astrophysics and
cosmology strictly speaking, understanding the equilibrium state
of self-gravitating systems is hence important also from a
broader, fundamental viewpoint \cite{gross174}.

In a recent work \cite{prl,epjb}, to be later referred to as I, we
investigated the possible equilibrium shapes of three-dimensional
rotating self-gravitating gases using a microcanonical mean-field
theory in which the system is enclosed in a finite spherical
volume $V$ with the total energy $E$ and the total angular
momentum $\boldsymbol{L}$ conserved. The singularity of the
Newtonian potential at short distances was cured by the use of
Lynden-Bell statistics \cite{lyndenbell67}. It turned out that,
along with the known ``gaseous'' and ``collapsed'' (a single,
dense cluster) states, the system can form symmetric ``binaries''
(two identical clusters) if $E$ is sufficiently low, that is if
gravity is strong enough, and rotation is sufficiently fast.
Moreover, at intermediate $E$ and $\boldsymbol{L}$, a large
phase-coexistence region with negative specific heat appears,
where the system is found in dense structures (either a single
cluster or a binary) embedded in a vapor.

The purpose of this paper is to complete the study of I by
describing the formation and thermodynamics of asymmetric binaries
(which are far more common in the universe than symmetric ones).
Furthermore, a family of order parameters will be introduced to
discern the possible equilibria and allow for a classification of
the equilibrium states. Some ``exotic'' equilibria (with a very
low entropy) that can be obtained using these order parameters are
presented as an example.

We begin by recalling the key steps in the theory of I. The
Hamiltonian is ($i,j=1,\ldots,N$)
\begin{equation}
H=\frac{1}{2m}\sum_i
p_i^2-Gm^2\sum_{i<j}|\boldsymbol{r}_i-\boldsymbol{r}_j|^{-1}
\end{equation}
with $\boldsymbol{r}_i\in V\subset\mathbb{R}^3$ denoting the
position and $\boldsymbol{p}_i\in\mathbb{R}^3$ the momentum of the
$i$-th particle, $m$ being its mass. $G$ is the gravitational
constant. The task is to find the particle density profiles that
maximize the Boltzmann entropy $S=k\log W$, where the state sum
$W$ is given by
\begin{equation}
W=\frac{\alpha}{N!}\int \delta(H-E)
~\delta(\boldsymbol{L}-\sum_i\boldsymbol{r}_i
\times\boldsymbol{p}_i)~d^{3N}\boldsymbol{r}~d^{3N}\boldsymbol{p}
\end{equation}
with $\alpha$ a constant. $W$ is a function of $E$ and
$\boldsymbol{L}$, as are $S$ and the desired density profiles. The
radius $R$ of the box is taken to be fixed. Adhering to the
notation of I, we shift to the reduced variable
$\boldsymbol{x}=\boldsymbol{r}/R$, which we shall indicate as
$\boldsymbol{x}=(x_1,x_2,x_3)$ (Cartesian representation) or
$\boldsymbol{x}=(x,\theta,\phi)$ (spherical representation).
Particles are assumed to have ``hard cores'', so that the density
profile $c$ is normalized as $\int
c(\boldsymbol{x})d\boldsymbol{x}=\Theta$, with $0\leq c\leq 1$.
$\Theta$ is a fixed parameter ($0\leq\Theta\leq 4\pi/3$) which
throughout this study, as in I, is taken to be $\Theta=0.02$. A
brief discussion about the dependence of the results on $\Theta$
is given at the end of the paper. Finally, we take $m=1$ and
measure energy and angular momentum in units of $GN^2/R$ and
$\sqrt{RGN^3}$, respectively.

Assuming Lynden-Bell statistics to mimic the hard cores
\cite{lyndenbell67}, at stationary points of the entropy surface
the density profile $c$ satisfies the condition (see I for
details)
\begin{equation}\label{key}
\log\frac{c(\boldsymbol{x})}{1-c(\boldsymbol{x})}=
\frac{\beta}{\Theta}\int\frac{c(\boldsymbol{x'})}{
|\boldsymbol{x}-\boldsymbol{x'}|}~d\boldsymbol{x'}+\frac{1}{2}\beta(
\boldsymbol{\omega}\times\boldsymbol{x})^2-\mu
\end{equation}
where
\begin{equation}
\beta=\frac{3/2}{[E-\frac{1}{2}\boldsymbol{L}^T
\mathbb{I}^{-1}\boldsymbol{L}-\Phi[c]]}~,
\end{equation}
$\boldsymbol{\omega}=\mathbb{I}^{-1}\boldsymbol{L}$ is the angular
velocity, and $\mu$ is a Lagrange multiplier implementing the
constraint on $\Theta$. $\mathbb{I}$ stands for the inertia
tensor, with elements ($a,b=1,2,3$)
\begin{equation}\label{inert}
I_{ab}[c]=\frac{1}{\Theta}\int
c(\boldsymbol{x})(x^2\delta_{ab}-x_a x_b)d\boldsymbol{x}
\end{equation}
(in units of $NR^2$), while $\Phi$ is the gravitational potential.
In general, the solutions of Eq. (\ref{key}), whose number and
form depend on $E$ and $\boldsymbol{L}$, can be written in
spherical coordinates as
\begin{equation}\label{rho}
c(\boldsymbol{x})=\sum_{l=0}^\infty \sum_{m=-l}^l
b_{lm}(x)Y_{lm}(\theta,\phi)
\end{equation}
where the ``weights'' $b_{lm}(x)$ have to be determined by solving
(numerically) the system of integral equations
\begin{gather}
b_{lm}(x)=\int g(x,\theta,\phi) Y_{lm}(\theta,\phi)~d\cos\theta
~d\phi\label{due}\\ g(x,\theta,\phi)=\l[1+e^{\frac{\beta}{\Theta}
\sum_{l,m} u_{lm}(x)Y_{lm}(\theta,\phi) -\frac{1}{2}\beta\omega^2
x^2 \sin^2\theta+\mu}\r]^{-1}\nonumber
\end{gather}
Some details of this task are discussed in I.

If $\boldsymbol{L}$ is taken to lie along the $3$-axis and if only
even harmonics ($l=0,2,4,\ldots$) are included in the series
(\ref{rho}), entropy-maximizing solutions of (\ref{key}) can be
divided in two main classes (see I): (a) axial-rotationally
symmetric ones, e.g. gas-cloud (high $E$, low $L$), distorted gas
clouds or disks (high $E$, high $L$), and single clusters (low
$E$, low $L$); (b) axial-rotationally asymmetric ones, like
identical double clusters or symmetric binaries (SBs, low $E$,
high $L$). The former are independent of the $\phi$ angle, and
thus for them the series (\ref{rho}) involves only the
\emph{zonal} harmonics, with $m=0$. The latter, instead, must
depend on $\phi$ and the lowest-order term in (\ref{rho}) that can
provide this dependence is the \emph{sectoral} harmonics
$Y_{2,2}$. In fact, on physical grounds, SBs can be characterized
by a non-zero value of the order parameter
\begin{equation}\label{d22}
D_{2,2}=\l|\int x^4 ~b_{2,2}(x)dx\r|
\end{equation}
which is proportional to the difference between the diagonal
components of the inertia tensor in the $1$ and $2$ directions.
Using (\ref{inert}) and (\ref{rho}) it is indeed easy to see that
\begin{equation}\label{d22inert}
D_{2,2}=\Theta\sqrt{\frac{15}{16\pi}}~|I_{11}-I_{22}|
\end{equation}

\begin{figure}[!]
\subfigure[$E=-0.60$,
$L=0.46$]{\scalebox{.38}{\includegraphics{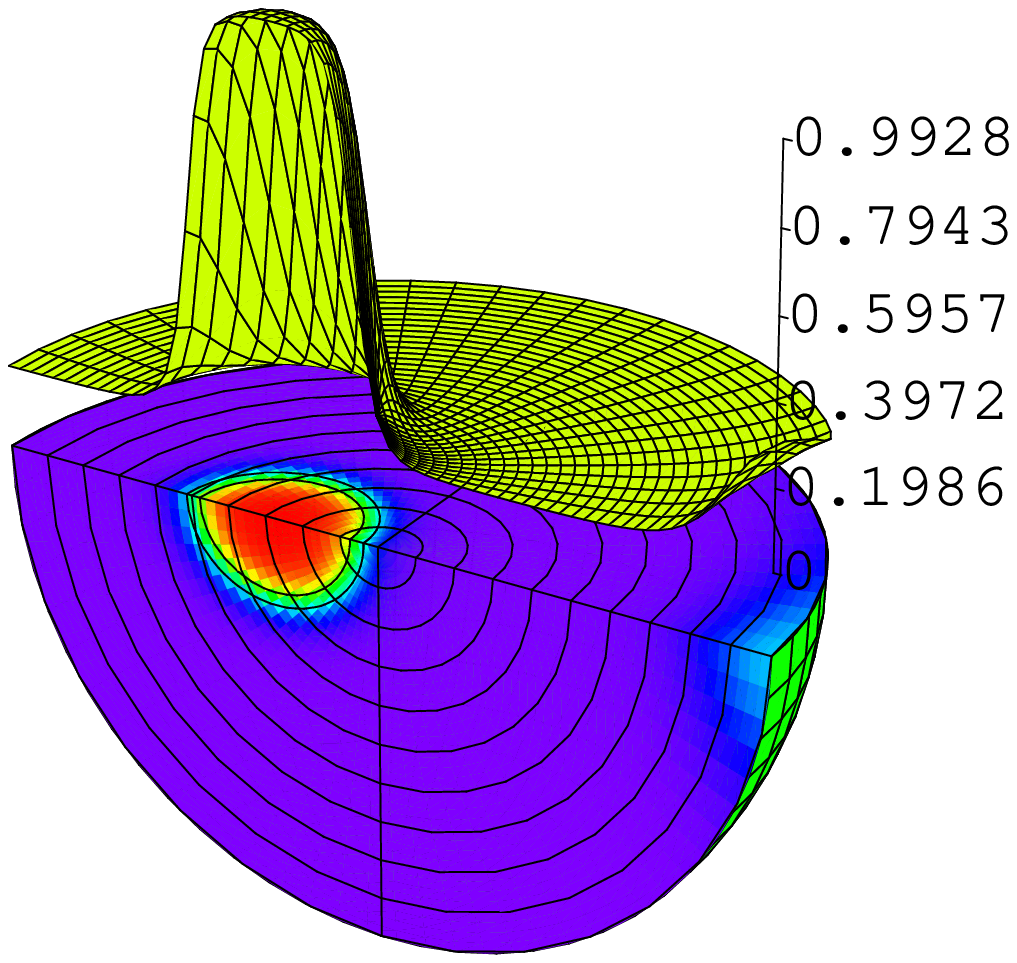}}}
\subfigure[$E=-1.16$,
$L=0.46$]{\scalebox{.42}{\includegraphics{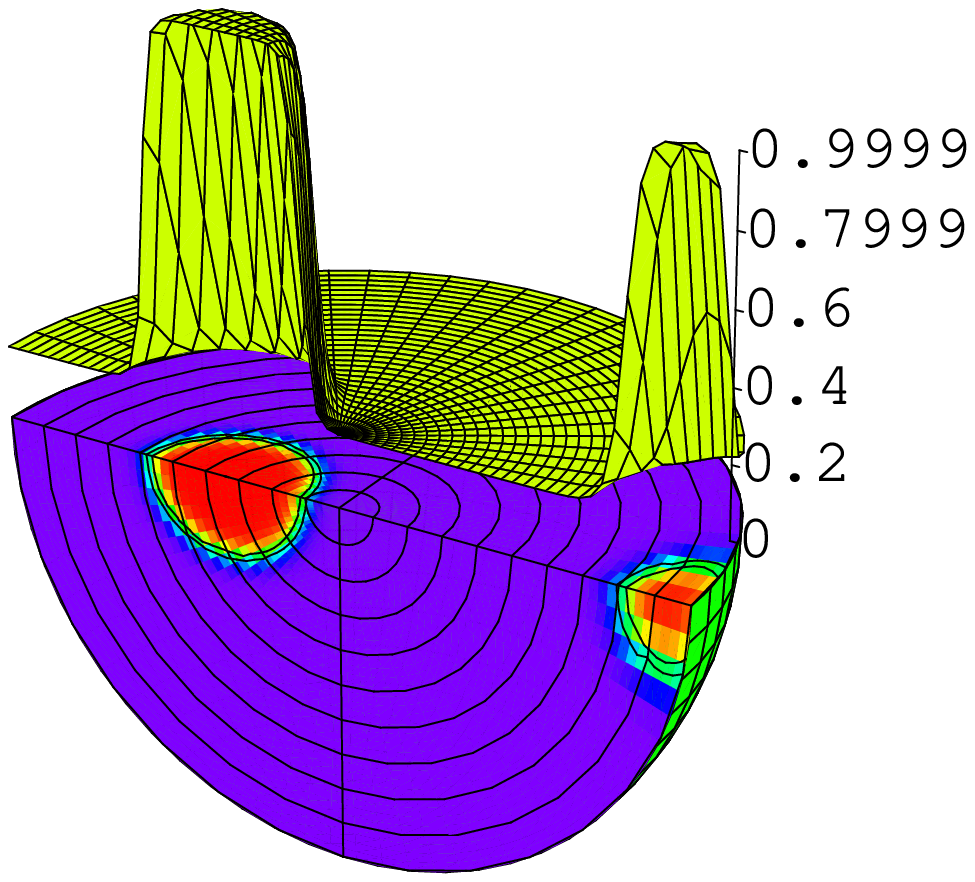}}}
\caption{\label{binaries} Two asymmetric binaries obtained from
Eq. (\ref{key}). The contour plot is shown together with the
density profile. (a) Incipient binary, with a broad low-density
body detached from the central cluster. (b) Well-formed binary
(very low energy).}
\end{figure}

\begin{figure}[b]
\includegraphics[height=8.5cm,angle=-90,clip=true]{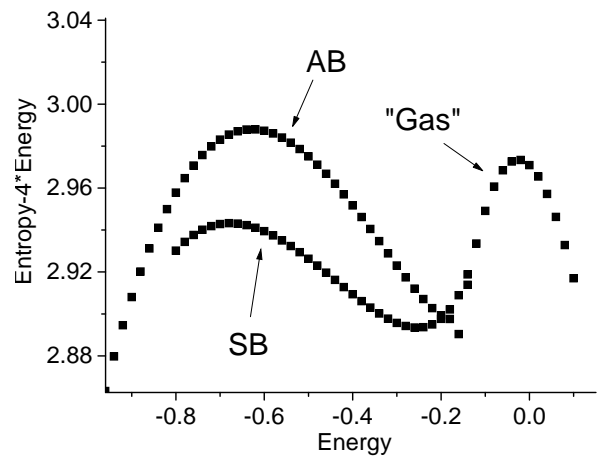}
\caption{\label{entropyL046}Entropy of the different stable
solutions of Eq. (\ref{key}) at $L=0.46$. The quantity $4E$ has
been subtracted from the entropy to make the separation and the
curvature more evident.}
\end{figure}

The inclusion of odd harmonics leads to new solutions, namely
asymmetric binaries (ABs). A sample of asymmetric binaries
obtained in this way (with the addition of fixing the center of
mass at $\boldsymbol{x}=\boldsymbol{0}$) is shown in Fig.
\ref{binaries}.
As it was to be expected, such solutions appear at low energies at
intermediate angular momenta ($L\simeq 0.46$) and exist in a small
range of values of $L$ up to $L\simeq 0.5$. This whole region lies
in the mixed phase with negative specific heat of I, hence
asymmetric binaries at equilibrium are embedded in a vapor and
they don't exist as a ``pure'' phase. Moreover, in this range they
are significantly more probable than symmetric binaries, as shown
in Fig. \ref{entropyL046}, where the entropy of the different
stable solutions at $L=0.46$ is shown.

A simple physical argument suffices to explain why at high angular
momenta ABs must have a much lower entropy than SBs, if they exist
at all. SBs have a larger moment of inertia than ABs, hence a
smaller rotational energy. As a consequence, for SBs the
``random'' kinetic contribution $E-L^2/(2I)$ is larger than for
ABs. Thus they have a larger entropy. So ABs are expected to be
strongly suppressed or significantly less probable at high angular
momenta. Indeed, we were unable to find stable asymmetric binaries
for $L>0.5$. The fact that ABs tend to form the smaller cluster in
proximity of the boundary of the box is due to the fact that ABs
start to form by evaporating small masses from a large, dense
cluster. The conservation of the center of mass relegates them to
the border. In well formed binaries, the peak of density in the
smaller cluster is separated from the boundary.

\begin{figure}
\includegraphics[height=8.5cm,angle=-90,clip=true]{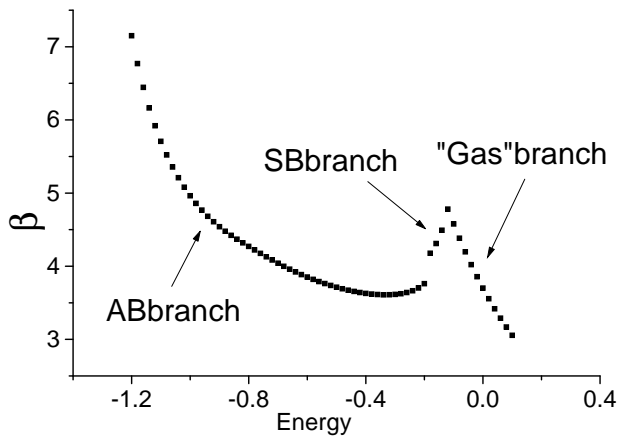}
\caption{\label{betaL046}Caloric curve $\beta$ vs $E$ at $L=0.46$.
Notice that the curvature of the entropy surface as a function of
$E$ and $L$, i.e. $S_{EE}S_{LL}-S_{EL}^2$, is negative along the
$AB$ branch.}
\end{figure}

In Fig. \ref{betaL046} the caloric curve $\beta=\partial
S/\partial E$ versus $E$ is shown at $L=0.46$.
The jump occurring at $E=-0.2$, where ABs become more favorable
than SBs, is an artifact of the asymptotics employed in the
mean-field theory of I. (Being $S$ a smooth and
multiply-differentiable function, such a feature is clearly
spurious.)

\begin{figure}[b]
\includegraphics[height=8.5cm,angle=-90,clip=true]{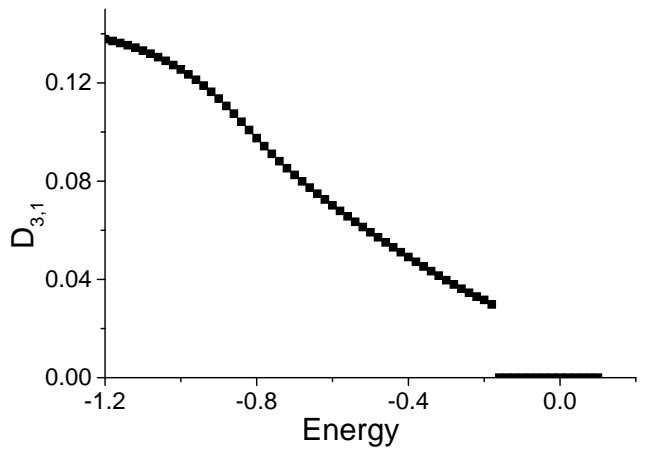}
\caption{\label{op}Behaviour of the order parameter for asymmetric
binaries $D_{3,1}$ as a function of energy at $L=0.46$.}
\end{figure}

Again on physical grounds, the natural order parameter for AB
solutions is $\avg{x_1^3}:=\int
x_1^3~c(\boldsymbol{x})d\boldsymbol{x}$, because in spatially
asymmetric structures it must be non-zero because of evident
symmetry reasons (the two clusters are aligned along the $x_1$
axis), at odds with what must happen in symmetric structures. Now
using (\ref{rho}) one can see by working out explicitly the
integrals involving spherical harmonics \cite{morse} that
\begin{multline}
\avg{x_1^3}=\sqrt{\frac{12\pi}{25}}\int x^5
b_{1,1}(x)dx-\sqrt{\frac{6\pi}{175}}\int x^5
b_{3,1}(x)dx+\\+\sqrt{\frac{2\pi}{35}}\int x^5 b_{3,3}(x)dx
\end{multline}
In principle, all of the above terms should be considered.
However, in any spatially symmetric structure (gas, single
cluster, SBs) all of them are zero, because odd harmonics do not
contribute, as discussed above. Each of them is non-zero only in
presence of spatial asymmetries, as in ABs, hence any of them can
be used as an order parameter. In particular, we find it
convenient to choose the quantity
\begin{equation}\label{d31}
D_{3,1}=\l|\int x^5 ~b_{3,1}(x)dx\r|
\end{equation}
whose behaviour at $L=0.46$ is shown in Fig. \ref{op}. AB
solutions appear at $E=-0.18$, where the order parameter jumps
discontinuously from zero to a finite value, and becomes more
favorable than SBs at $E=-0.2$.

Eq.s (\ref{d22}) and (\ref{d31}) suggest that, in general, one can
classify the different equilibrium states by introducing the
family of order parameters defined by
\begin{equation}\label{ops}
D_{l,m}=\l|\int x^{l+2}b_{l,m}(x)dx\r|
\end{equation}
with integers $l=0,1,2,\ldots$ and $m=-l,-l+1,\ldots,l$. For small
$l$, each of these functions has a simple physical interpretation.

For $l=0$, $D_{0,0}$ is simply a constant proportional to
$\Theta$, since, recalling that $\int
c(\boldsymbol{x})d\boldsymbol{x}\equiv\avg{1}=\Theta$,
\begin{equation}
\int c(\boldsymbol{x})d\boldsymbol{x}=2\sqrt{\pi}\int x^2
b_{0,0}(x) dx
\end{equation}

For $l=1$, one has that $D_{1,1}$, $D_{1,0}$ and $D_{1,-1}$ are
nothing but the coordinates of the center of mass (modulo a
proportionality constant):
\begin{gather}
x_1^{CM}\equiv\avg{x_1}=\sqrt{\frac{4\pi}{3}}\int x^3
b_{1,1}(x)dx\nonumber\\
x_2^{CM}\equiv\avg{x_2}=\sqrt{\frac{4\pi}{3}}\int x^3
b_{1,-1}(x)dx\label{cm}\\
x_3^{CM}\equiv\avg{x_3}=\sqrt{\frac{4\pi}{3}}\int x^3
b_{1,0}(x)dx\nonumber
\end{gather}

For $l=2$, the situation is slightly more involved. $D_{2,0}$ can
be easily seen to be linked to the quantity
\begin{equation}
\avg{x^2-3x_3^2}=-\sqrt{\frac{16\pi}{5}}\int x^4 b_{2,0}(x)dx
\end{equation}
which is a convenient order parameter for disk (deformed gas)
solutions. In fact, for an hypothetical configuration being a
perfect sphere (say a homogeneous gas cloud filling the whole
$V$), one would have $D_{2,0}=0$ evidently. When rotation deforms
the cloud by shrinking it along the $3$-axis, one would have
$D_{2,0}\neq 0$. As discussed in I, $D_{2,2}$ is instead the order
parameter discerning axial-rotationally symmetric structures from
axial-rotationally asymmetric ones (it is non zero for both SBs
and ABs). It is particularly convenient for SBs because of its
connection to the inertia tensor components, see (\ref{d22inert}).
As for the remaining $D_{2,m}$'s, they can be easily seen to be
related to the off-diagonal components of the inertia tensor
(\ref{inert}).
%\begin{gather}
%\avg{x_1 x_2}=\sqrt{\frac{4\pi}{15}}\int x^4 f_{2,-2}(x) dx\\
%\avg{x_1 x_3}=\sqrt{\frac{4\pi}{15}}\int x^4 f_{2,1}(x) dx\\
%\avg{x_2 x_3}=\sqrt{\frac{4\pi}{15}}\int x^4 f_{2,-1}(x) dx
%\end{gather}

It is clear that higher-order $D$'s are all connected to
homogeneous polynomials in $x_1$, $x_2$, $x_3$, which are in turn
connected to the Cartesian representation of spherical harmonics
\cite{morse}. For example, for $l=3$, $D_{3,0}$ is connected to
\begin{equation}
\avg{2x_3^3-3x_1^2x_3-3x_2^2x_3}=\sqrt{\frac{16\pi}{7}}\int x^5
b_{3,0}(x)dx
\end{equation}
Their physical interpretation for larger $l$ is however less
straightforward.

\begin{figure}[!]
\subfigure{\scalebox{.48}{\includegraphics{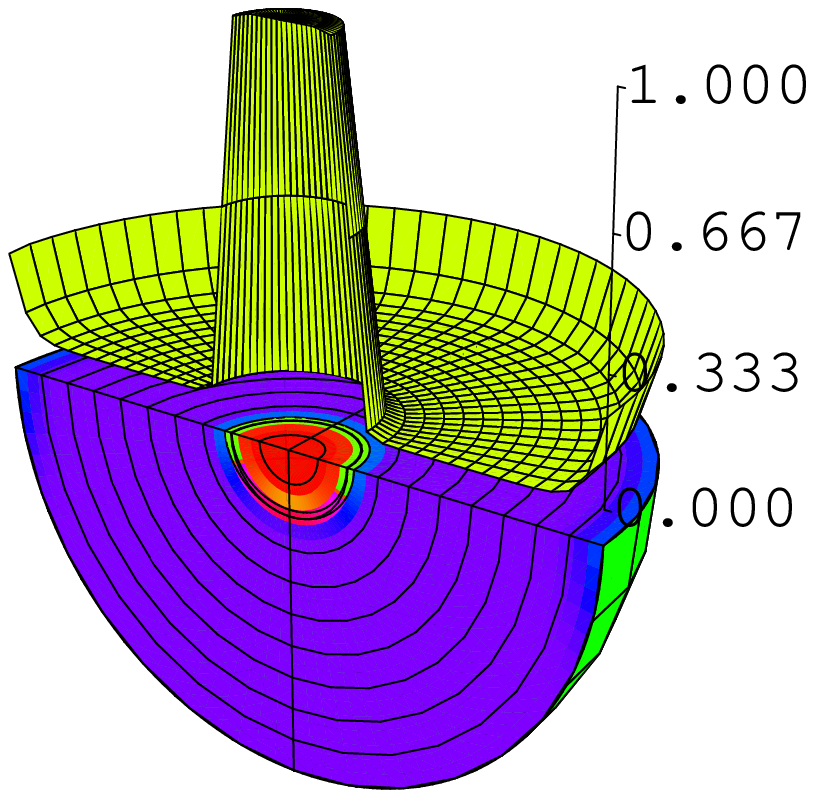}}}
\subfigure{\scalebox{.48}{\includegraphics{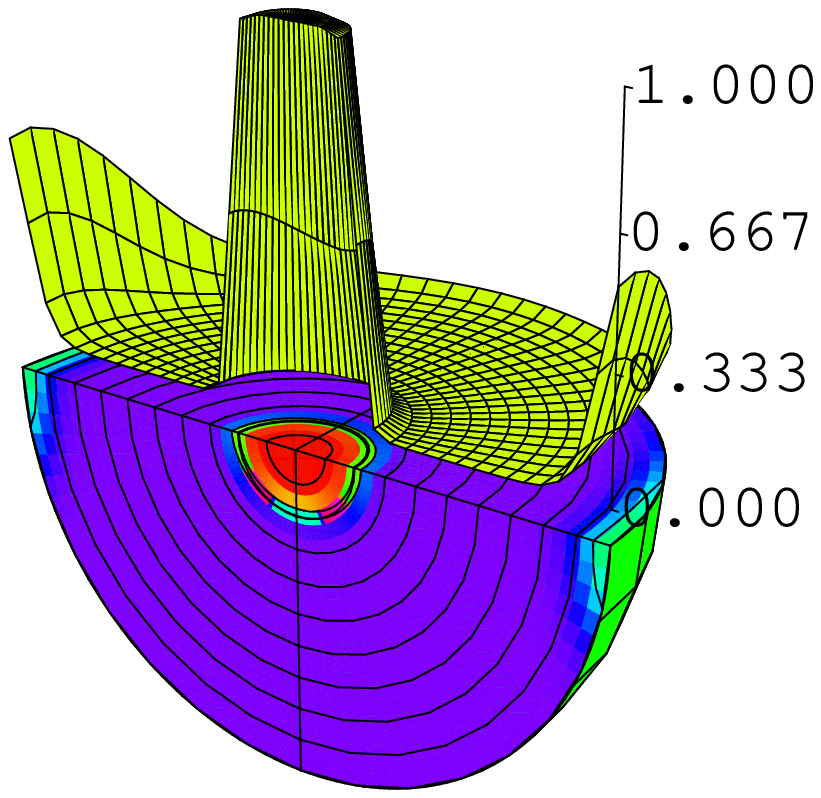}}}
\caption{\label{zoo} Two ``exotic'' solutions of (\ref{key})
obtained via the order parameters (\ref{ops}). (a) Single cluster
surrounded by a ring. (b) Single cluster with two ``satellites''.
These solutions coexist at $E=-0.5$ and $L=0.7$.}
\end{figure}

Using these order parameters, solutions can be classified
hierarchically, a procedure that is particularly effective at high
$L$. In our setting, with the center of mass fixed at
$\boldsymbol{x}=\boldsymbol{0}$ one has the following picture.
Homogeneous gas clouds have all $D_{lm}$'s equal to zero except
$D_{0,0}$. For deformed clouds (disks) only $D_{l,0}$ with $l$
even are non zero. For symmetric binaries (which are connected to
``gas''-like solutions continuously, see I), also $D_{2,2}\neq 0$
(and in general all $D_{l,m}$ with $l$ and $m$ even), all other
$D$'s being zero, while for asymmetric binaries $D_{3,1}$ is also
not zero. This classification can be continued further, for the
order parameters (\ref{op}) allow to construct all possible
solutions of (\ref{key}), by adding a proper field that selects
the solution for which the chosen order parameter is not zero. As
an example, we show in Fig. \ref{zoo} two of several equilibria
that can be obtained in this way at very high angular momentum. Of
course, such ``exotic'' solutions of (\ref{key}), if stable, have
much lower entropy than those discussed in I and here.

To summarize, we have carried further the equilibrium study of
rotating self-gravitating gases of classical non-overlapping
particles started in I by describing the formation and
thermodynamics of asymmetric binaries. Moreover, we introduced a
family of order parameters that allow to characterize all the
different shapes that can be encountered (homogeneous gas, disk,
symmetric binary, asymmetric binary, along with more exotic
states), and discussed their physical meaning in simple cases. The
present study does not substantially alter the global phase
diagram presented in I, since asymmetric binaries do not exist as
a pure thermodynamic state but only in the phase coexistence
region. Among the problems that remain open we would like to
mention the following two. First, the dependence on $\Theta$. Our
results have been obtained at fixed $\Theta=0.02$. Different
$\Theta$'s may lead to different results. By letting $\Theta\to 0$
one obtains the case of overlapping particles (hard cores are
neglected and Lynden-Bell statistics is replaced by Boltzmann
statistics in space), which is a classical problem in cosmology
dating back to \cite{antonov62}. The rotating case is dealt with
in \cite{npb}. For Boltzmann particles we haven't found any
evidence of asymmetric binaries. On the other hand, when $\Theta$
grows ($\Theta\leq 4\pi/3$) the system may become too closely
packed and the interesting features one observes for small
$\Theta$ might disappear at some point. The second problem is the
dependence on $R$. Throughout this paper we have assumed that $R$
is fixed but it would be interesting to analyze this problem in a
context in which $R$ varies, for instance to mimic an
expanding-universe scenario.

\end{document}